\documentclass[12pt]{article}
\pdfoutput=1

\usepackage[a4paper,text={16.8cm,22.4cm}]{geometry}
\usepackage{amsmath,amsfonts,amssymb,bm,graphicx,color,euscript}
\usepackage[small,labelfont=bf]{caption}

\RequirePackage[sort&compress,square,comma,numbers]{natbib}
\allowdisplaybreaks
\newcommand{\A}{{\EuScript A}}
\newcommand{\F}{{\EuScript F}}
\newcommand{\G}{{\EuScript G}}
\newcommand{\J}{{\EuScript J}}
\newcommand{\Ha}{{\EuScript H}}
\newcommand{\X}{{\EuScript X}}
\newcommand{\nsl}{\rlap{\hspace{0.25mm}/}{n}}
\newcommand{\Asl}{\rlap{\hspace{0.7mm}/}{\A}}
\newcommand{\Gsl}{\rlap{\hspace{0.2mm}/}{\G}}
\newcommand{\Dsl}{\rlap{\hspace{0.75mm}/}{D}}

\begin{document}

\begin{titlepage}

\begin{flushright}
\normalsize
MITP/20-011\\ 
March 6, 2020
\end{flushright}

\vspace{1.0cm}
\begin{center}
\Large\bf
{\boldmath Two-Loop Radiative Jet Function for Exclusive\\ 
$B$-Meson and Higgs Decays}
\end{center}

\vspace{0.5cm}
\begin{center}
Ze Long Liu$^a$ and Matthias Neubert$^{b,c}$\\
\vspace{0.7cm} 
{\sl ${}^a$Theoretical Division, Los Alamos National Laboratory, Los Alamos, NM 87545, U.S.A.\\[3mm]
${}^b$PRISMA$^+$ Cluster of Excellence \& Mainz Institute for Theoretical Physics\\
Johannes Gutenberg University, 55099 Mainz, Germany\\[3mm]
${}^c$Department of Physics \& LEPP, Cornell University, Ithaca, NY 14853, U.S.A.}
\end{center}

\vspace{0.8cm}
\begin{abstract}
The rare radiative $B$-meson decay $B^-\to\gamma\ell^-\bar\nu$ and the radiative Higgs-boson decay $h\to\gamma\gamma$ mediated by light-quark loops both receive large logarithmic corrections in QCD, which can be resummed using factorization theorems derived in soft-collinear effective theory. In these factorization theorems the same radiative jet function appears, which is a central object in the study of factorization beyond the leading order in scale ratios. We calculate this function at two-loop order both in momentum space and in a dual space, where its renormalization-group evolution takes on a simpler form. We also derive the two-loop anomalous dimension of the jet function and present the exact solution to its evolution equation at two-loop order. Another important outcome of our analysis is the explicit form of the two-loop anomalous dimension of the $B$-meson light-cone distribution amplitude in momentum space. 
\end{abstract}

\end{titlepage}

\section{Introduction}

Soft-collinear effective theory (SCET) is a convenient tool to study the factorization properties of cross sections and scattering or decay amplitudes sensitive to several hierarchical energy scales \cite{Bauer:2000yr,Bauer:2001yt,Bauer:2002nz,Beneke:2002ph,Beneke:2002ni}. Typically, the corresponding factorization theorems contain hard functions, jet functions and soft functions, which receive contributions from different momentum regions in Feynman diagrams. The hard functions correspond to Wilson coefficients obtained when the full theory is matched onto SCET, whereas the jet or soft functions are defined in terms of matrix elements of collinear or soft fields in the low-energy effective theory.

Jet functions -- matrix elements of non-local products of collinear fields -- play an important role in these factorization theorems. They often live at an intermediate scale, which lies between the hard-scattering scale of the process and lowest energy scale it is sensitive to. The most familiar jet function is defined as the spectral function (i.e., the discontinuity) of the quark propagator dressed by a light-like Wilson line connecting the two quark fields \cite{Becher:2006qw}. This jet function appears in a large variety of phenomenological applications. For example, it enters in the factorization theorems for inclusive $B$-meson decays into light final-state particles, such as $\bar B\to X_s\gamma$ and $\bar B\to X_u\,\ell^-\bar\nu$ \cite{Korchemsky:1994jb,Bosch:2004th}. It also appears in the resummation of threshold logarithms in deep inelastic scattering \cite{Becher:2006nr,Becher:2006mr}.

Recently, there has been a growing interest in understanding factorization at subleading power in scale ratios. Beyond the leading power a large variety of hard, collinear and soft functions appear. In particular, while at leading power soft emissions are eikonal and can be described by soft Wilson lines, at subleading power the emission of soft fermions and power-suppressed emissions of soft gauge bosons need to be taken into account. Particularly interesting is the case of soft quark emission, which is absent at leading power. At subleading order in the SCET expansion, there is a unique interaction that couples a soft quark to collinear quarks and gauge fields. In the notation of \cite{Beneke:2002ph}, it reads 
\begin{equation}\label{subleadingL}
   {\cal L}_{q\,\xi_n}^{(1/2)}(x)
   = \bar q_s(x_-)\,W_n^\dagger(x)\,i\Dsl_n^\perp\,\xi_n(x) + \mbox{h.c.} \,, 
\end{equation}
where $\xi_n$ is a collinear quark spinor subject to the constraint $\nsl\,\xi_n=0$, $W_n$ is a collinear Wilson line, $D_n^\mu$ is a covariant collinear derivative (containing collinear gauge fields) acting on collinear fields, and $q_s$ describes a soft quark. The collinear particles carry large momentum flow along a light-like direction $n^\mu$. The soft quark field must be multipole expanded for consistency, and we denote $x_-^\mu=(\bar n\cdot x)\,\frac{n^\mu}{2}$, where $\bar n^\mu$ is a conjugate light-like vector satisfying $n\cdot\bar n=2$ (see \cite{Becher:2014oda} for a pedagogical introduction to SCET).

New jet functions can be defined in terms of the matrix elements of collinear fields in the presence of one insertion of this subleading Lagrangian \cite{Beneke:2018gvs,Moult:2019mog,Beneke:2019oqx}. These functions are called ``radiative jet functions'' \cite{DelDuca:1990gz,Bonocore:2015esa,Bonocore:2016awd}, since they involve the emission of a soft particle from inside a jet. The radiative jet functions are of general interest not only because they appear in the description of power corrections to established factorization theorems. There exist interesting physical processes that are sensitive to soft quark exchange already at leading order in the expansion of the relevant decay or scattering amplitude. One of the first appearances of a radiative jet function $J(p^2)$ related to collinear interactions with a soft quark appeared in the theoretical description of the exclusive, radiative $B$-meson decay $B^-\to\gamma\ell^-\bar\nu$ \cite{Lunghi:2002ju,Bosch:2003fc}. In this process the soft spectator quark of the $B$ meson couples to a collinear photon and an off-shell collinear quark, which then connects to the weak-interaction vertex, where it annihilates the $b$-quark and turns into a virtual $W^-$ boson. Interestingly, the same radiative jet function has recently been encountered in a completely different context: in the theoretical description of the contribution to the radiative Higgs-boson decay $h\to\gamma\gamma$ that is induced by light $b$-quark loops \cite{Liu:2019oav,Wang:2019mym}. (This is not the dominant contribution to the decay amplitude, but it is a particularly interesting one with regard to its factorization properties.) 

Following our recent work \cite{Liu:2019oav}, we define the radiative jet function $J(p^2)$ in terms of the matrix element 
\begin{equation}\label{defJF}
   \int d^dx\,e^{ip_s\cdot x_-}\,\langle\gamma(k)|\,
    T\,\big( W_n^\dagger\,i\Dsl_n^\perp\,\xi_n \big)(x)\,
    \big( \bar\xi_n W_n \big)(0)\,|0\rangle 
   = e_q\,\rlap/\varepsilon_\perp^*(k)\,\frac{\nsl}{2}\,\frac{i\bar n\cdot k}{p^2+i0}\,J(p^2) \,, 
\end{equation}
where $p_s$ is the momentum carried away by the soft quark in (\ref{subleadingL}), and $p\equiv k+p_{s+}$. We denote by $e_q$ the electric charge of the collinear quark. Note that $p_s\cdot x_-=p_{s+}\cdot x$ with $p_{s+}^\mu=(n\cdot p_s)\,\frac{\bar n^\mu}{2}$. The jet function $J(p^2)$ depends on the only non-trivial kinematic invariant (note that $k^2=0$ and $p_s^2=0$)
\begin{equation}
   p^2\equiv (k+p_{s+})^2 = \bar n\cdot k\,n\cdot p_s \,.
\end{equation}

In Section~\ref{sec:2} we present a detailed discussion of some general properties of the radiative jet function $J(p^2)$, with a special focus on its behavior under renormalization-group (RG) evolution. We derive the two-loop anomalous dimension of the jet function and present exact solutions to its RG evolution equations both in momentum space and in the so-called dual space. The technique we employ for obtaining these solutions is general and can be applied to other radiative jet functions, too. Section~\ref{sec:3} contains a description of the calculation of the jet function at two-loop order in QCD. The renormalization of the jet function is discussed in Section~\ref{sec:4}. In Section~\ref{sec:5} we briefly comment on phenomenological implications of our results in the context of $B^-\to\gamma\ell^-\bar\nu$ decay. We then present our conclusions.

\section{General properties the radiative jet function}
\label{sec:2}

We begin by reviewing and deriving some general properties the jet function, some of which are based on insights that were uncovered a long time ago, while several others are new.

\subsection{One-loop expressions}

At one-loop order, we find that the bare jet function in $d=4-2\epsilon$ spacetime dimensions reads
\begin{equation}
   J^{(0)}(p^2) = 1 + \frac{C_F\alpha_{s,0}}{4\pi} \left( - p^2 - i0 \right)^{-\epsilon}
    e^{\epsilon\gamma_E}\,\frac{\Gamma(1+\epsilon)\,\Gamma^2(-\epsilon)}{\Gamma(2-2\epsilon)}\,
    (2-4\epsilon-\epsilon^2) + {\cal O}(\alpha_{s,0}^2) \,.
\end{equation}
Renormalizing the bare coupling in the $\overline{\rm MS}$ scheme,
\begin{equation}
   \alpha_{s,0} = \mu^{2\epsilon}\,Z_\alpha\,\alpha_s(\mu) \,, \qquad
   Z_\alpha = 1 - \beta_0\,\frac{\alpha_s}{4\pi\epsilon} + {\cal O}(\alpha_s^2) \,,
\end{equation}
where $\beta_0=\frac{11}{3}\,C_A-\frac43\,T_F\,n_f$ is the first coefficient of the QCD $\beta$-function, one obtains
\begin{equation}
   J^{(0)}(p^2) = 1 + \frac{C_F\alpha_s}{4\pi} \left( \frac{-p^2-i0}{\mu^2} \right)^{-\epsilon}
    \left[ \frac{2}{\epsilon^2} - 1 - \frac{\pi^2}{6} + {\cal O}(\epsilon) \right] 
    + {\cal O}(\alpha_s^2) \,.
\end{equation}
Here and below $\alpha_s\equiv\alpha_s(\mu)$ always denotes the renormalized coupling. For simplicity, we will from now on drop the ``$-i0$'' prescription, which defines the sign of the imaginary part of the jet function in the time-like region, where $p^2>0$.

While at one-loop order one could renormalize the jet function by means of a local counterterm, the correct renormalization factor has a more complicated non-local form.\footnote{It is an embarrassment that there is no known method in SCET to derive the anomalous dimensions of jet functions directly from their operator definitions.} 
The proper renormalization condition has been derived from the consistency of the factorization formula for the $B^-\to\gamma\ell^-\bar\nu$ decay amplitude, requiring that the amplitude be independent of the renormalization scale \cite{Bosch:2003fc}. In this process, the known RG equations for the $B$-meson light-cone distribution amplitude (LCDA) \cite{Lange:2003ff} and some other quantities have been used. The jet function depends on a single argument $p^2$, which can be either time-like or space-like. The time-like (space-like) jet functions belonging to different $p^2>0$ ($p^2<0$) values mix under renormalization, but there is no mixing between the time-like and space-like jet functions. For the time-like case, we write the renormalization condition in the form
\begin{equation}
   J(p^2,\mu) = \frac{1}{p^2} \int_0^\infty\!dp^{\prime\,2}\,
    Z_J(p^2,p^{\prime\,2};\mu)\,J^{(0)}(p^{\prime\,2}) \,,
\end{equation}
with a dimensionless renormalization factor $Z_J$. 
For the space-like case an analogous expression holds, where $p^{\prime\,2}$ is integrated over the interval $(-\infty,0\hspace{0.3mm}]$. Treating both cases at the same time, we write the renormalization condition in the form 
\begin{equation}\label{Jrenorm}
   J(p^2,\mu) = \int_0^\infty\!dx\,Z_J(p^2,x p^2;\mu)\,J^{(0)}(x p^2) \,.
\end{equation}
At one-loop order, one finds (the generalization with $y\ne 1$ is needed below)
\begin{equation}\label{ZJ}
   Z_J(y p^2,x p^2;\mu) 
   = \left[ 1 + \frac{C_F\alpha_s}{4\pi} \left( - \frac{2}{\epsilon^2} 
    + \frac{2}{\epsilon} \ln\frac{-p^2}{\mu^2} \right) \right] \delta(y-x) 
    + \frac{C_F\alpha_s}{2\pi\epsilon}\,\Gamma(y,x)
    + {\cal O}(\alpha_s^2) \,,
\end{equation}
where the symmetric distribution
\begin{equation}
   \Gamma(y,x) = \left[ \frac{\theta(y-x)}{y(y-x)} + \frac{\theta(x-y)}{x(x-y)} \right]_+
\end{equation}
arises in the so-called Lange-Neubert kernel for the $B$-meson LCDA \cite{Lange:2003ff} (see also \cite{Grozin:1996pq}). The plus prescription is defined such that, when $\Gamma(y,x)$ is integrated with a function $f(x)$, one must replace $f(x)\to f(x)-f(y)$ under the integral. At one-loop order the plus distribution has no effect when the renormalized jet function is derived from (\ref{Jrenorm}). One finds
\begin{equation}
   J(p^2,\mu) = 1 + \frac{C_F\alpha_s}{4\pi}
    \left[ \ln^2\!\bigg(\frac{-p^2}{\mu^2}\bigg) - 1 - \frac{\pi^2}{6} \right] 
    + {\cal O}(\alpha_s^2) \,.
\end{equation}
This result was first obtained in \cite{Lunghi:2002ju,Bosch:2003fc}. One of the main goals of this paper is to calculate the two-loop corrections to this formula.

\subsection{Renormalization-group evolution}
\label{subsec:2.2}

The renormalized jet function obeys the RG evolution equation
\begin{equation}\label{RGE}
   \frac{d}{d\ln\mu}\,J(p^2,\mu) 
   = - \int_0^\infty\!dx\,\gamma_J(p^2,x p^2;\mu)\,J(x p^2,\mu) \,,
\end{equation}
where the anomalous dimension is defined as
\begin{equation}\label{neweq}
   \gamma_J(p^2,x p^2;\mu)
   = - \int_0^\infty\!dy\,\frac{dZ_J(p^2,y p^2;\mu)}{d\ln\mu}\,Z_J^{-1}(y p^2,x p^2;\mu) \,.
\end{equation}
As usual, it can be obtained from the coefficient of the single $1/\epsilon$ pole in $Z_J$ via \cite{Becher:2005pd}
\begin{equation}
   \gamma_J(p^2,x p^2;\mu) 
   = 2\alpha_s\,\frac{\partial Z_J^{[1]}(p^2,x p^2;\mu)}{\partial\alpha_s} \,.
\end{equation}
Without loss of generality, we express the result in the form \cite{Bosch:2003fc}
\begin{equation}\label{gammaJ}
   \gamma_J(p^2,x p^2;\mu) 
   = \left[ \Gamma_{\rm cusp}(\alpha_s)\,\ln\frac{-p^2}{\mu^2}
    - \gamma'(\alpha_s) \right] \delta(1-x) + \Gamma_{\rm cusp}(\alpha_s)\,\Gamma(1,x) 
    + {\cal O}(\alpha_s^2) \,,
\end{equation}
where we have identified the coefficient of the logarithmic term with the light-like cusp anomalous dimension $\Gamma_{\rm cusp}(\alpha_s)$ in the fundamental representation of SU$(N_c)$, a central quantity in the theory of the renormalization of Wilson loops with cusps \cite{Korchemsky:1987wg,Korchemskaya:1992je}. Since the plus distribution is linked with the logarithmic term, it is multiplied by the same quantity. The non-logarithmic term $\gamma'(\alpha_s)$ of the local part of the anomalous dimension vanishes at one-loop order \cite{Bosch:2003fc}. Its two-loop expression will be derived here for the first time. 

The first three terms in the anomalous dimension $\gamma_J$ retain their form to all orders in perturbation theory. The most general form of the local terms is a linear function of $\ln(-p^2/\mu^2)$, and the cusp anomalous dimension is the coefficient of the logarithmic term. This is the only term depending on the momentum variable $p^2$ alone. For dimensional reasons, all remaining terms can only depend on the dimensionless ratio $x=p^{\prime\,2}/p^2$. The form of the non-local terms (with $x\ne 1$) is presently only known at one-loop order. The ${\cal O}(\alpha_s^2)$ corrections indicated in (\ref{gammaJ}) thus refer to higher-order non-local terms, which are presently unknown. For the more familiar jet function entering inclusive processes such as the rare inclusive decay $\bar B\to X_s\gamma$, such higher-order corrections are known to be absent \cite{Becher:2006qw}, i.e.\ the functional form of the one-loop anomalous dimension is preserved in higher orders, and the non-local terms are determined completely by the cusp anomalous dimension. We will see, however, that further non-local higher-order terms {\em do exist\/} in the case of the exclusive jet function in (\ref{defJF}). 

For the remainder of this section, we will ignore the unknown higher-order non-local terms in (\ref{gammaJ}), but we will keep the remaining terms at arbitrary order in perturbation theory. It is then possible to derive exact solutions to the RG evolution equation (\ref{RGE}) using a technique developed in \cite{Bosch:2003fc,Lange:2003ff}. It is based on the observation that on dimensional grounds the integral (the variables $x$ and $y$ can carry arbitrary but equal mass dimension)
\begin{equation}
   \F(a) \equiv \int_0^\infty\!dx\,y\,\Gamma(y,x) \left( \frac{x}{y} \right)^a
   = - \big[ \psi(1+a) + \psi(1-a) + 2\gamma_E \big] 
\end{equation}
evaluates to a dimensionless function of the exponent $a$. Here $\psi(z)=\Gamma'(z)/\Gamma(z)$ is the digamma function. It can then be checked that the ansatz
\begin{equation}
   \left( \frac{-p^2}{\mu_j^2} \right)^{\eta+a_\Gamma(\mu_j,\mu)}
    \exp\Bigg[ - 2 S(\mu_j,\mu) - a_{\gamma'}(\mu_j,\mu) 
    - \int\limits_{\alpha_s(\mu_j)}^{\alpha_s(\mu)}\!d\alpha\,
    \frac{\Gamma_{\rm cusp}(\alpha)}{\beta(\alpha)}\,
    \F\big(\eta+a_\Gamma(\mu_j,\mu_\alpha)\big) \Bigg] 
\end{equation}
with $\alpha_s(\mu_\alpha)\equiv\alpha$ provides a solution to the RG equation with the initial condition $(-p^2/\mu_j^2)^\eta$ at some matching scale $\mu=\mu_j$, at which $J(p^2,\mu_j)$ is assumed to be free of large logarithms. Here $\beta(\alpha_s)=d\alpha_s(\mu)/d\ln\mu$ is the QCD $\beta$-function, and we have defined the RG functions \begin{equation}
   S(\mu_j,\mu) 
   = - \int\limits_{\alpha_s(\mu_j)}^{\alpha_s(\mu)}\!
    d\alpha\,\frac{\Gamma_{\rm cusp}(\alpha)}{\beta(\alpha)}
    \int\limits_{\alpha_s(\mu_j)}^\alpha
    \frac{d\alpha'}{\beta(\alpha')} \,, \qquad
   a_\Gamma(\mu_j,\mu) 
   = - \int\limits_{\alpha_s(\mu_j)}^{\alpha_s(\mu)}\!
    d\alpha\,\frac{\Gamma_{\rm cusp}(\alpha)}{\beta(\alpha)} \,,
\end{equation}
which are the solutions to the equations 
\begin{equation}
   \frac{d}{d\ln\mu}\,S(\mu_j,\mu) = - \Gamma_{\rm cusp}(\alpha_s)\,\ln\frac{\mu}{\mu_j} \,,
    \qquad
   \frac{d}{d\ln\mu}\,a_\Gamma(\mu_j,\mu) = - \Gamma_{\rm cusp}(\alpha_s) \,. 
\end{equation}
The function $a_{\gamma'}(\mu_j,\mu)$ is defined analogously to $a_\Gamma(\mu_j,\mu)$. Note that both $S(\mu_j,\mu)$ and $a_\Gamma(\mu_j,\mu)$ take negative (positive) values if $\mu>\mu_j$ ($\mu<\mu_j$), since the cusp anomalous dimension is positive. Explicit expressions for these objects obtained at next-to-next-to-leading order (NNLO) in perturbation theory can be found in the appendix of \cite{Becher:2006mr}. 

At any fixed order in perturbation theory, the renormalized jet function at the matching scale depends on $p^2$ only via powers of the logarithm $L_p=\ln(-p^2/\mu_j^2)$. We can generate these logarithms by taking derivatives with respect to $\eta$. Hence, with the definition 
\begin{equation}\label{eq44}
   J(p^2,\mu_j) \equiv \J(L_p,\mu_j)
\end{equation}
the exact solution of the evolution equation can be written in the closed form \cite{Bosch:2003fc}
\begin{equation}
\begin{aligned}
   J(p^2,\mu)
   &= \J(\partial_\eta,\mu_j) \left( \frac{-p^2}{\mu_j^2} \right)^{\eta+a_\Gamma(\mu_j,\mu)} 
    \exp\Big[ - 2 S(\mu_j,\mu) - a_{\gamma'}(\mu_j,\mu) \Big] \\
   &\quad\times
    \exp\Bigg[ - \int\limits_{\alpha_s(\mu_j)}^{\alpha_s(\mu)}\!d\alpha\,
    \frac{\Gamma_{\rm cusp}(\alpha)}{\beta(\alpha)}\,
    \F\big(\eta+a_\Gamma(\mu_j,\mu_\alpha)\big) \Bigg] \Bigg|_{\eta=0} \,.
\end{aligned}
\end{equation}
Here and below, derivatives with respect to the auxiliary parameter $\eta$ always act on all terms standing to the right. We now change integration variables in the exponent of the last term from $\alpha$ to $\varrho=a_\Gamma(\mu_j,\mu_\alpha)$. This yields
\begin{equation}
   - \int\limits_{\alpha_s(\mu_j)}^{\alpha_s(\mu)}\!d\alpha\,
    \frac{\Gamma_{\rm cusp}(\alpha)}{\beta(\alpha)}\,
    \F\big(\eta+a_\Gamma(\mu_j,\mu_\alpha)\big) 
   = \int\limits_{0}^{a}\!d\varrho\,\,\F(\eta+\varrho) 
   = \ln\frac{\Gamma\big(1-\eta-a\big)\,\Gamma(1+\eta)}{\Gamma\big(1+\eta+a\big)\,\Gamma(1-\eta)} 
    - 2\gamma_E\,a \,,
\end{equation}
where $a\equiv a_\Gamma(\mu_j,\mu)$. This leads to the final result \cite{Bosch:2003fc}
\begin{equation}\label{Jsolelegant}
\begin{aligned}
   J(p^2,\mu)
   &= \exp\Big[ - 2 S(\mu_j,\mu) - a_{\gamma'}(\mu_j,\mu) - 2\gamma_E\,a_\Gamma(\mu_j,\mu) \Big] \\
   &\quad\times \J(\partial_\eta,\mu_j) 
    \left( \frac{-p^2}{\mu_j^2} \right)^{\eta+a_\Gamma(\mu_j,\mu)} 
    \frac{\Gamma\big(1-\eta-a_\Gamma(\mu_j,\mu)\big)\,\Gamma(1+\eta)}%
         {\Gamma\big(1+\eta+a_\Gamma(\mu_j,\mu)\big)\,\Gamma(1-\eta)}\,\bigg|_{\eta=0} \,.
\end{aligned}
\end{equation}

An alternative solution of the RG equation (\ref{RGE}), which works for more general initial conditions (even though this is not needed for the case at hand), can be obtained by taking a Fourier transform of the jet function with respect to $\ln(-p^2/\mu^2)$, i.e.\
\begin{equation}
   J(p^2,\mu) = \frac{1}{2\pi}\,\int_{-\infty}^\infty\!dt\,\tilde J(t,\mu)
    \left( \frac{-p^2}{\mu^2} \right)^{it} .
\end{equation}
Since the right-hand side of this equation exhibits a power-like dependence on $(-p^2/\mu^2)$, one can use the technique described above (with $\eta$ replaced by $it$) to show that for general initial condition the evolution equation (\ref{RGE}) is solved by 
\begin{equation}\label{Jsol}
\begin{aligned}
   J(p^2,\mu) 
   &= \exp\Big[ -2S(\mu_j,\mu) - a_{\gamma'}(\mu_j,\mu) - 2\gamma_E\,a_\Gamma(\mu_j,\mu) \Big] \\
   &\quad\times \frac{1}{2\pi}\,\int_{-\infty}^\infty\!dt\,\tilde J(t,\mu_j)
    \left( \frac{-p^2}{\mu_j^2} \right)^{it+a_\Gamma(\mu_j,\mu)}
    \frac{\Gamma\big(1-it-a_\Gamma(\mu_j,\mu)\big)\,\Gamma(1+it)}%
         {\Gamma\big(1+it+a_\Gamma(\mu_j,\mu)\big)\,\Gamma(1-it)} \,.
\end{aligned}
\end{equation}
This solution forms the basis of the construction of the jet function in the so-called dual space (see below). For the case of downward scale evolution, for which $\mu<\mu_j$, it is possible to evaluate the integral over $t$ in closed form. This is discussed in Appendix~\ref{app:C}.

\subsection{Jet function in the dual space}

For the case of the $B$-meson LCDA, it has been shown in \cite{Bell:2013tfa} that one can bring an RG equation with an anomalous dimension of the type shown in (\ref{gammaJ}) -- in the approximation where unknown, non-local contributions to the anomalous dimension arising at two-loop order and higher are neglected -- to a much simpler form using a suitably chosen integral transform. Adapted to our case, the key observation based on (\ref{Jsol}) is that the function 
\begin{equation}
   g(t,\mu) 
   \equiv \frac{\Gamma(1+it)}{\Gamma(1-it)}\,\tilde J(t,\mu) \left( \frac{-p^2}{\mu^2} \right)^{it} 
\end{equation}
has a particulary simple behavior under RG evolution. Shifting the integration variable in (\ref{Jsol}) from $t$ to $t'=t-ia_\Gamma(\mu_j,\mu)$, one finds that
\begin{equation}
   g(t,\mu) 
   = \left( \frac{-p^2 e^{-2\gamma_E}}{\mu_j^2} \right)^{a_\Gamma(\mu_j,\mu)}
    \exp\Big[ -2S(\mu_j,\mu) - a_{\gamma'}(\mu_j,\mu) \Big]\,
    g\big(t+ia_\Gamma(\mu_j,\mu),\mu_j\big) \,.
\end{equation}
Defining a dual jet function $j(p^2,\mu)$ via the Fourier transform
\begin{equation}\label{jdual}
   j(p^2,\mu) = \frac{1}{2\pi} \int_{-\infty}^\infty\!dt\,\frac{\Gamma(1+it)}{\Gamma(1-it)}\,
    \tilde J(t,\mu) \left( \frac{-p^2}{\mu^2} \right)^{it} ,
\end{equation}
we then obtain
\begin{equation}
   j(p^2,\mu) 
   = \left( \frac{-p^2 e^{-2\gamma_E}}{\mu_j^2} \right)^{a_\Gamma(\mu_j,\mu)}
    \exp\Big[ -2S(\mu_j,\mu) - a_{\gamma'}(\mu_j,\mu) \Big]\,j(p^2,\mu_j) \,.
\end{equation}
This dual function obeys the {\em local\/} RG equation
\begin{equation}\label{localRGE}
   \frac{d}{d\ln\mu}\,j(p^2,\mu) 
   = - \left[ \Gamma_{\rm cusp}(\alpha_s)\,\ln\frac{-p^2 e^{-2\gamma_E}}{\mu^2}
    - \gamma'(\alpha_s) \right] j(p^2,\mu) \,.
\end{equation}
We stress again that this equation only holds in the approximation where the unknown non-local contributions to the anomalous dimension (\ref{gammaJ}) are neglected. Equation (\ref{2lJevol}) below shows the generalization required when these terms are included at two-loop order.

The relation between the original function and the dual function can be derived by combining (\ref{jdual}) and (\ref{inverseFT}). This gives
\begin{equation}
   j(p^2,\mu) = \int_0^\infty\!\frac{dx}{x}\,J(x p^2,\mu)\,
    \frac{1}{2\pi} \int_{-\infty}^\infty\!dt\,\frac{\Gamma(1+it)}{\Gamma(1-it)}\,x^{-it} \,.
\end{equation}
The integrand of the $t$-integral has poles at values $t=in$ with $n\in\mathbb{N}$. Evaluating the integral using the theorem of residues, one obtains 
\begin{equation}
   \frac{1}{2\pi} \int_{-\infty}^\infty\!dt\,\frac{\Gamma(1+it)}{\Gamma(1-it)}\,x^{-it} 
   = \sqrt{x}\,\,{\rm J}_1\big(2\sqrt{x}\big) \,,
\end{equation}
where ${\rm J}_1(x)$ is a Bessel function. One thus finds the integral transforms \cite{Bell:2013tfa}
\begin{equation}\label{trafo}
\begin{aligned}
   j(p^2,\mu) 
   &= \int_0^\infty\!\frac{dx}{\sqrt{x}}\,\,{\rm J}_1\big(2\sqrt{x}\big)\,J(x p^2,\mu) \,, \\
   J(p^2,\mu) 
   &= \int_0^\infty\!\frac{dx}{\sqrt{x}}\,\,{\rm J}_1\big(2\sqrt{x}\big)\,j(p^2/x,\mu) \,,
\end{aligned}
\end{equation}
where the second relation follows from the orthonormality condition 
\begin{equation}\label{orthonorm}
   \int_0^\infty\!dx\,\,{\rm J}_1\big(2\sqrt{xa}\big)\,{\rm J}_1\big(2\sqrt{xb}\big) = \delta(a-b) \,.
\end{equation}
At one-loop order, we find that
\begin{equation}
   j(p^2,\mu) = 1 + \frac{C_F\alpha_s}{4\pi}
    \left[ \ln^2\!\bigg(\frac{-p^2 e^{-2\gamma_E}}{\mu^2}\bigg) - 1 - \frac{\pi^2}{6} \right] 
    + {\cal O}(\alpha_s^2) \,.
\end{equation}

In the beautiful papers \cite{Braun:2014owa,Braun:2018fiz} it was shown that the Lange-Neubert kernel for the $B$-meson LCDA can be written in a remarkably compact form as a logarithm of the generator of special conformal transformations along the light-cone. Using tools from conformal field theory, the above-mentioned transformation of the evolution equation (\ref{RGE}) to the local form (\ref{localRGE}) was rederived. In subsequent work by the same authors \cite{Braun:2016qlg,Braun:2019wyx} the evolution equation was extended to two-loop order. It was found that, starting at ${\cal O}(\alpha_s^2)$, non-local terms appear in the dual space as well. We will use these results in our two-loop analysis below.

\subsection{Two-loop evolution of the jet function}

The two-loop RG equation obeyed by the jet function in the dual space can be derived from the QCD factorization theorem for the decay $B^-\to\gamma\ell^-\bar\nu$ valid in the region of large photon energy ($E_\gamma\lesssim m_b/2$). At leading power in $\Lambda_{\rm QCD}/m_b$, the corresponding decay amplitude can be written in the factorized form \cite{Lunghi:2002ju,Bosch:2003fc}
\begin{equation}\label{QCDF}
   {\cal M}(B^-\to\gamma\ell^-\bar\nu)
   \propto F_B(\mu)\,H(m_b,2E_\gamma,\mu)\,\int_0^\infty\!\frac{d\omega}{\omega}\,
    J(-2E_\gamma\omega,\mu)\,\phi_+^B(\omega,\mu) \,,
\end{equation}
where $F_B$ is related to the $B$-meson decay constant in the heavy-quark limit ($F_B\approx f_B\sqrt{m_B}$ modulo radiative corrections), $H$ is a hard-scattering function depending on the short-distance scales $m_b$ and $2E_\gamma$, and $\phi_+^B$ denotes the leading-twist LCDA of the $B$ meson depending on a variable $\omega={\cal O}(\Lambda_{\rm QCD})$ \cite{Grozin:1996pq}. $J$ is the jet function discussed above, with $p^2<0$ in the space-like region. This function depends on an intermediate scale of order $2E_\gamma\omega\sim m_b\Lambda_{\rm QCD}$. In the dual space, the right-hand side of (\ref{QCDF}) takes an identical form \cite{Bell:2013tfa}, i.e.\
\begin{equation}
   {\cal M}(B^-\to\gamma\ell^-\bar\nu)
   \propto F_B(\mu)\,H(m_b,2E_\gamma,\mu)\,\int_0^\infty\!\frac{d\omega}{\omega}\,
    j(-2E_\gamma\omega,\mu)\,\rho_+(\omega,\mu) \,,
\end{equation}
where the dual function $j$ is related to the original jet function by the first relation in (\ref{trafo}), and $\rho_+$ is defined in an analogous way as
\begin{equation}
   \rho_+(\omega,\mu) 
   = \int_0^\infty\!\frac{dx}{\sqrt{x}}\,\,{\rm J}_1\big(2\sqrt{x}\big)\,\phi_+^B(x\omega,\mu) \,.  
\end{equation}

In \cite{Braun:2019wyx}, the RG evolution equation for the function $\eta_+(s,\mu)$, which is related to $\rho_+(\omega,\mu)$ via $s\,\eta_+(s,\mu)=\rho_+(1/s,\mu)$, was derived at two-loop order. Interestingly, it was observed that at this order the evolution of the dual function is no longer local. Rather, it was found that (we present the equation for $\rho_+$ rather than $\eta_+$)
\begin{equation}\label{eq37}
\begin{aligned}
   \frac{d}{d\ln\mu}\,\rho_+(\omega,\mu) 
   &= \left[ \Gamma_{\rm cusp}(\alpha_s)\,\ln\frac{\omega\,e^{-2\gamma_E}}{\mu}
    - \gamma_\eta(\alpha_s) \right] \rho_+(\omega,\mu) \\
   &\quad\mbox{}+ C_F \left( \frac{\alpha_s}{2\pi} \right)^2 
    \int_0^1\!\frac{dx}{1-x}\,h(x)\,\rho_+(\omega/x,\mu) + {\cal O}(\alpha_s^3) \,,
\end{aligned}
\end{equation}
where 
\begin{equation}\label{hdef}
   h(x) = \ln x \left[ \beta_0 
    + 2C_F \left( \ln x - \frac{1+x}{x}\,\ln(1-x) - \frac32 \right) \right]
\end{equation}
arises from conformal symmetry breaking. We can now use the known RG equations for the $B$-meson decay constant in heavy-quark effective theory \cite{Broadhurst:1991fz}
\begin{equation}
   \frac{d}{d\ln\mu}\,F_B(\mu) = - \gamma_F(\alpha_s)\,F_B(\mu)
\end{equation}
and of the hard-scattering function \cite{Neubert:2004dd,Asatrian:2008uk,Bell:2008ws,Becher:2009kw}
\begin{equation}
   \frac{d}{d\ln\mu}\,H(m_b,2E_\gamma,\mu) 
   = \left[ \Gamma_{\rm cusp}(\alpha_s)\,\ln\frac{2E_\gamma}{\mu} + \gamma_H(\alpha_s) \right]  
    H(m_b,2E_\gamma,\mu)
\end{equation}
to derive the two-loop evolution equation for the jet function in the dual space. We obtain 
\begin{equation}\label{2lJevol}
\begin{aligned}
   \frac{d}{d\ln\mu}\,j(p^2,\mu) 
   &= - \left[ \Gamma_{\rm cusp}(\alpha_s)\,\ln\frac{-p^2 e^{-2\gamma_E}}{\mu^2}
    - \gamma'(\alpha_s) \right] j(p^2,\mu) \\
   &\quad\mbox{}- C_F \left( \frac{\alpha_s}{2\pi} \right)^2 
    \int_0^1\!\frac{dx}{1-x}\,h(x)\,j(x p^2,\mu) + {\cal O}(\alpha_s^3) \,,
\end{aligned}
\end{equation}
where
\begin{equation}\label{gammaprela}
   \gamma'(\alpha_s) = \gamma_\eta(\alpha_s) - \gamma_H(\alpha_s) + \gamma_F(\alpha_s) \,. 
\end{equation}
The two-loop expressions for the anomalous dimensions on the right-hand side of this relation are listed in Appendix~\ref{app:B}. Using these results, we obtain
\begin{equation}\label{gprela}
   \gamma'(\alpha_s) 
   = C_F \left( \frac{\alpha_s}{4\pi} \right)^2 
    \left[ C_A \bigg( \frac{808}{27} - \frac{11\pi^2}{9} - 28\zeta_3 \bigg) 
    - T_F\,n_f \bigg( \frac{224}{27} - \frac{4\pi^2}{9} \bigg) \right]
   + {\cal O}(\alpha_s^3) \,.
\end{equation}
This quantity is genuinely non-abelian; it starts at two-loop order and has no $C_F^2$ term. Interestingly, we find that $\gamma'(\alpha_s)=-\gamma_W(\alpha_s)$ coincides, up to a sign, with the anomalous dimension $\gamma_W$ of the Drell-Yan soft function derived in \cite{Becher:2007ty}. The same quantity is known to arise in the evolution equations for the thrust \cite{Becher:2008cf}, beam thrust \cite{Stewart:2010qs} and hemisphere soft functions \cite{Hornig:2011iu} and for the soft function appearing in transverse-momentum resummation \cite{Li:2016ctv}. It would be interesting to explore the nature of this connection in more detail. 

Starting from (\ref{2lJevol}), we can apply the second transformation rule in (\ref{trafo}) to derive the explicit form of the RG evolution equation (\ref{RGE}) for the jet function in momentum space at two-loop order. The relevant anomalous dimension $\gamma_J$ is given by
\begin{equation}\label{magicrelation}
   \gamma_J(p^2,xp^2,\mu)
   = \int_0^\infty\!\frac{dy}{\sqrt{xy}} \int_0^\infty\!dz\,\,{\rm J}_1\big(2\sqrt{z}\big)\,
    {\rm J}_1\bigg(2\sqrt{\frac{xz}{y}}\bigg)\,\gamma_J^{\rm dual}(p^2/z,yp^2/z,\mu) \,,
\end{equation}
where $\gamma_J^{\rm dual}$ denotes the anomalous dimension in the dual space, which according to (\ref{2lJevol}) is given by
\begin{equation}\label{gammaJ2loop}
\begin{aligned}
   \gamma_J^{\rm dual}(p^2,x p^2;\mu)
   &= \left[ \Gamma_{\rm cusp}(\alpha_s)\,\ln\frac{-p^2 e^{-2\gamma_E}}{\mu^2}
    - \gamma'(\alpha_s) \right] \delta(1-x) \\
   &\quad\mbox{}+ C_F \left( \frac{\alpha_s}{2\pi} \right)^2 
    \frac{\theta(1-x)}{1-x}\,h(x) + {\cal O}(\alpha_s^3) \,.
\end{aligned}
\end{equation}
Using the orthonormality condition (\ref{orthonorm}), we obtain from (\ref{magicrelation})
\begin{equation}
\begin{aligned}
   \gamma_J(p^2,x p^2;\mu) 
   &= \left[ \Gamma_{\rm cusp}(\alpha_s)\,\ln\frac{-p^2 e^{-2\gamma_E}}{\mu^2}
    - \gamma'(\alpha_s) \right] \delta(1-x) \\
   &\quad\mbox{}- \Gamma_{\rm cusp}(\alpha_s)\,\frac{1}{\sqrt x}
    \int_0^\infty\!dz\,\ln z\,\,{\rm J}_1\big(2\sqrt{z}\big)\,{\rm J}_1\big(2\sqrt{xz}\big) \\
   &\quad\mbox{}+ C_F \left( \frac{\alpha_s}{2\pi} \right)^2 \frac{\theta(1-x)}{1-x}\,h(x) 
    + {\cal O}(\alpha_s^3) \,.
\end{aligned}
\end{equation}
The integral in the second line diverges for $x\to 1$ and must be evaluated in the sense of distributions by studying its action on a smooth test function $f(x)$. We find
\begin{equation}
\begin{aligned}
   \frac{1}{\sqrt x} \int_0^\infty\!dz\,\ln z\,\,
    {\rm J}_1\big(2\sqrt{z}\big)\,{\rm J}_1\big(2\sqrt{xz}\big) 
   = - \left[ \frac{\theta(1-x)}{1-x} + \frac{\theta(x-1)}{x(x-1)} \right]_+
    - 2\gamma_E\,\delta(1-x) \,.
\end{aligned}
\end{equation}
Using this result, we obtain 
\begin{equation}\label{RGEmom}
\begin{aligned}
   \gamma_J(p^2,x p^2;\mu) 
   &= \left[ \Gamma_{\rm cusp}(\alpha_s)\,\ln\frac{-p^2}{\mu^2}
    - \gamma'(\alpha_s) \right] \delta(1-x) + \Gamma_{\rm cusp}(\alpha_s)\,\Gamma(1,x) \\
   &\quad\mbox{}+ C_F \left( \frac{\alpha_s}{2\pi} \right)^2 \frac{\theta(1-x)}{1-x}\,h(x) 
    + {\cal O}(\alpha_s^3) \,.
\end{aligned}
\end{equation}
This is the desired extension of relation (\ref{gammaJ}) to two-loop order.

As an important tangential outcome of our analysis, we now derive the explicit form of the two-loop RG evolution equation of the $B$-meson LCDA in momentum space. Because of the structural similarity of the evolution equations (\ref{eq37}) and (\ref{2lJevol}) in the dual space, we can apply the same method as above to obtain
\begin{equation}
\begin{aligned}
   \frac{d}{d\ln\mu}\,\phi_+^B(\omega,\mu) 
   &= \left[ \Gamma_{\rm cusp}(\alpha_s)\,\ln\frac{\omega}{\mu}
    - \gamma_\eta(\alpha_s) \right] \phi_+^B(\omega,\mu) 
    + \Gamma_{\rm cusp}(\alpha_s) \int_0^\infty\!dx\,\Gamma(1,x)\,\phi_+^B(\omega/x,\mu) \\
   &\quad\mbox{}+ C_F \left( \frac{\alpha_s}{2\pi} \right)^2 
    \int_0^1\!\frac{dx}{1-x}\,h(x)\,\phi_+^B(\omega/x,\mu) + {\cal O}(\alpha_s^3) \,.
\end{aligned}
\end{equation}
This is the desired extension of the Lange-Neubert kernel to two-loop order.

\subsection{Solutions to the two-loop evolution equations}

The evolution equation (\ref{2lJevol}) in the dual space can be solved using the same technique we have adopted in Section~\ref{subsec:2.2}. The key observation is that any power $\left(-p^2\right)^a$ of the momentum squared is an eigenfunction of the evolution kernel. To see this, note that 
\begin{equation}
   \int_0^1\!\frac{dx}{1-x}\,h(x)\,x^a = \beta_0\,\Ha(a)
\end{equation}
defines a dimensionless function of the exponent $a$, where
\begin{equation}
   \Ha(a) = \left( \frac{3C_F}{\beta_0} -1 \right) \psi'(1+a) 
    + \frac{2C_F}{\beta_0} \left[ \frac{\psi'(1+a)}{a}
    - \left( \frac{1}{a^2} + 2\psi'(1+a) \right) \big( \psi(1+a) + \gamma_E \big) \right] .
\end{equation}
It follows that the function 
\begin{equation}
\begin{aligned}
   &\left( \frac{-p^2 e^{-2\gamma_E}}{\mu_j^2} \right)^{\eta+a_\Gamma(\mu_j,\mu)}
    \exp\Big[ - 2 S(\mu_j,\mu) - a_{\gamma'}(\mu_j,\mu) \Big] \\
   &\times \exp\Bigg[
    - \int\limits_{\alpha_s(\mu_j)}^{\alpha_s(\mu)}\!\!
    \frac{d\alpha}{\beta(\alpha)}\,\bigg[ C_F \left( \frac{\alpha}{2\pi} \right)^2 
    \beta_0\,\Ha\big(\eta+a_\Gamma(\mu_j,\mu_\alpha)\big) + {\cal O}(\alpha^3) \bigg] \Bigg] 
\end{aligned}
\end{equation}
provides a solution to (\ref{2lJevol}) with the initial condition $(-p^2 e^{-2\gamma_E}/\mu_j^2)^\eta$ at the matching scale $\mu=\mu_j$. We now use the fact that the initial condition $j(p^2,\mu_j)$ depends on $p^2$ only through powers of the logarithm $\hat L_p=\ln(-p^2 e^{-2\gamma_E}/\mu^2)$. Writing 
\begin{equation}\label{eq52}
   j(p^2,\mu_j) \equiv \hat\J(\hat L_p,\mu_j) \,,
\end{equation}
we find in analogy with (\ref{Jsolelegant}) that the general solution to the evolution equation (\ref{2lJevol}) in the dual space is obtained as
\begin{equation}\label{wundervoll1}
\begin{aligned}
   j(p^2,\mu)
   &= \exp\Big[ - 2 S(\mu_j,\mu) - a_{\gamma'}(\mu_j,\mu) \Big] \\
   &\quad\times \hat\J(\partial_\eta,\mu_j)\! 
    \left(\! \frac{-p^2 e^{-2\gamma_E}}{\mu_j^2} \!\right)^{\eta+a_\Gamma(\mu_j,\mu)} 
    \exp\!\Bigg[ C_F\!\!\int\limits_{\alpha_s(\mu_j)}^{\alpha_s(\mu)}\!
    \frac{d\alpha}{2\pi}\,\Big[ \Ha\big(\eta+a_\Gamma(\mu_j,\mu_\alpha)\big) 
    + {\cal O}(\alpha) \Big] \Bigg] \Bigg|_{\eta=0} ,
\end{aligned}
\end{equation}
where we have used that $\beta(\alpha_s)=-\beta_0\alpha_s^2/(2\pi)+\dots$ at leading order.

It is not difficult to transform this solution back to momentum space. Using the second relation in (\ref{trafo}), we find that
\begin{equation}
\begin{aligned}
   J(p^2,\mu)
   &= \exp\Big[ - 2 S(\mu_j,\mu) - a_{\gamma'}(\mu_j,\mu) \Big] \\
   &\quad\times \hat\J(\partial_\eta,\mu_j) 
    \left( \frac{-p^2 e^{-2\gamma_E}}{\mu_j^2} \right)^{\eta+a_\Gamma(\mu_j,\mu)} 
    \frac{\Gamma\big(1-\eta-a_\Gamma(\mu_j,\mu)\big)}%
         {\Gamma\big(1+\eta+a_\Gamma(\mu_j,\mu)\big)} \\
   &\quad\times \exp\Bigg[ C_F\!\!\int\limits_{\alpha_s(\mu_j)}^{\alpha_s(\mu)}\!
    \frac{d\alpha}{2\pi}\,\Big[ \Ha\big(\eta+a_\Gamma(\mu_j,\mu_\alpha)\big) 
    + {\cal O}(\alpha) \Big] \Bigg] \Bigg|_{\eta=0} \,.
\end{aligned}
\end{equation}
In the final step, we employ the relation
\begin{equation}
   \J(\partial_\eta,\mu_j)
   = \hat\J(\partial_\eta,\mu_j)\,e^{-2\gamma_E\eta}\,\frac{\Gamma(1-\eta)}{\Gamma(1+\eta)}\,
    \bigg|_{\eta=0}
   = \hat\J(\partial_\eta,\mu_j) \left[ 1 + \frac{2\zeta_3}{3}\,\eta^3 + {\cal O}(\eta^5) \right] 
    \bigg|_{\eta=0}
\end{equation}
between the functions $\hat\J(\partial_\eta,\mu_j)$ defined in (\ref{eq52}) and $\J(\partial_\eta,\mu_j)$ defined in (\ref{eq44}), which follows by setting $\mu=\mu_j$ in the above solution. We conclude that the general solution to the momentum-space RG evolution equation (\ref{RGE}) is given by
\begin{equation}\label{wundervoll2}
\begin{aligned}
   J(p^2,\mu)
   &= \exp\bigg[ - 2 S(\mu_j,\mu) - a_{\gamma'}(\mu_j,\mu) 
    - 2\gamma_E\,a_\Gamma(\mu_j,\mu) \bigg] \\
   &\quad\times \J(\partial_\eta,\mu_j) 
    \left( \frac{-p^2}{\mu_j^2} \right)^{\eta+a_\Gamma(\mu_j,\mu)} 
    \frac{\Gamma\big(1-\eta-a_\Gamma(\mu_j,\mu)\big)\,\Gamma(1+\eta)}%
         {\Gamma\big(1+\eta+a_\Gamma(\mu_j,\mu)\big)\,\Gamma(1-\eta)} \\
   &\quad\times \exp\Bigg[ C_F\!\!\int\limits_{\alpha_s(\mu_j)}^{\alpha_s(\mu)}\!
    \frac{d\alpha}{2\pi}\,\Big[ \Ha\big(\eta+a_\Gamma(\mu_j,\mu_\alpha)\big) 
    + {\cal O}(\alpha) \Big] \Bigg] \Bigg|_{\eta=0} \,.
\end{aligned}
\end{equation}

The explicit expressions for the two-loop anomalous dimensions (\ref{gammaJ2loop}) and (\ref{RGEmom}), as well as the explicit solutions of the corresponding evolution equations (\ref{wundervoll1}) and (\ref{wundervoll2}), are among the most important new results obtained in this paper.

\section{Two-loop calculation of the bare jet function}
\label{sec:3}

The radiative jet function can be directly evaluated from its definition in (\ref{defJF}). In doing so, we work to leading order in the electromagnetic coupling $e$ but include higher-order QCD corrections. It is useful to recast the original definition in terms of so-called ``gauge-invariant collinear building blocks'' defined as \cite{Bauer:2002nz,Hill:2002vw}
\begin{equation}
   \X_n(x) = W_n^\dagger(x)\,\xi_n(x) \,, \qquad
   \A_n^\mu(x) + \G_n^\mu(x) = W_n^\dagger(x) \big[ iD_n^\mu\,W_n(x) \big] \,.
\end{equation}
Note that the effective photon field $\A_n^\mu$ and the gluon field $\G_n^\mu$, which contain the gauge couplings in their definition, are separately gauge invariant. The definition of the jet function in (\ref{defJF}) therefore defines two gauge-invariant objects $J_A(p^2)$ and $J_G(p^2)$ via
\begin{equation}\label{JAJGdef}
\begin{aligned}
   &\int d^dx\,e^{ip_s\cdot x_-} \langle\gamma(k)|\,
    T\,\big( \Asl_n^\perp(x) + \Gsl_n^\perp(x) \big)\,\X_n(x)\,\bar\X_n(0)\,|0\rangle \\
   &= e_q\,\rlap/\varepsilon_\perp^*(k)\,\frac{\nsl}{2}\,\frac{i\bar n\cdot k}{p^2+i0}\,
    \big[ J_A(p^2) + J_G(p^2) \big] \,. 
\end{aligned}
\end{equation}
A third contribution involving the derivative $\rlap/\partial_\perp$ acting on $\X_n(x)$ vanishes, because this field carries vanishing transverse momentum. 

Since we work to leading order in electromagnetic interactions, the function $J_A$ is equivalent to the collinear quark propagator dressed by light-like Wilson lines. This object was studied in \cite{Becher:2006qw}, where its discontinuity at two-loop order was presented. The function $J_A$ can be determined from this result in a straightforward way. The two-loop calculation of the new jet function $J_G$ can be performed using similar methods. As shown in \cite{Becher:2006qw}, the propogator in (\ref{defJF}) can be rewritten in terms of standard QCD fields, because the collinear SCET Lagrangian (without couplings to soft fields) is equivalent to the original QCD Lagrangian \cite{Beneke:2002ph}. We therefore use QCD Feynman rules for convenience. 

In principle, the matrix element for $J_G$ can be calculated in a general covariant gauge. However, the Feynman rules for vertices derived from the collinear gluon field ${\cal G}^\mu(x)$ are rather complicated due to the Wilson lines contained in its definition. In light-cone gauge $n\cdot A(x)=0$, on the other hand, the field ${\cal G}^\mu(x)=g_s A^\mu(x)$ takes on a very simple form, since the Wilson lines become trivial ($W_n=1$). The smaller number of Feynman diagrams and the absence of ghost contributions result in a more efficient computation of $J_G$ in this gauge. The free gluon propagator with momentum $l^\mu$ in light-cone gauge is given by 
\begin{equation}
   \frac{i}{l^2+i0} \left( - g^{\mu\nu} + \frac{{\bar n}^\mu l^\nu
    + {\bar n}^\nu l^\mu}{\bar n\cdot l} \right) ,
\end{equation}
where we do not adopt the Mandelstam-Leibbrandt prescription to regularize the singularity at ${\bar n}\cdot l=0$  (see \cite{Becher:2010pd} for more details). 

\begin{figure}[t]
\begin{center}
\includegraphics[width=0.85\textwidth]{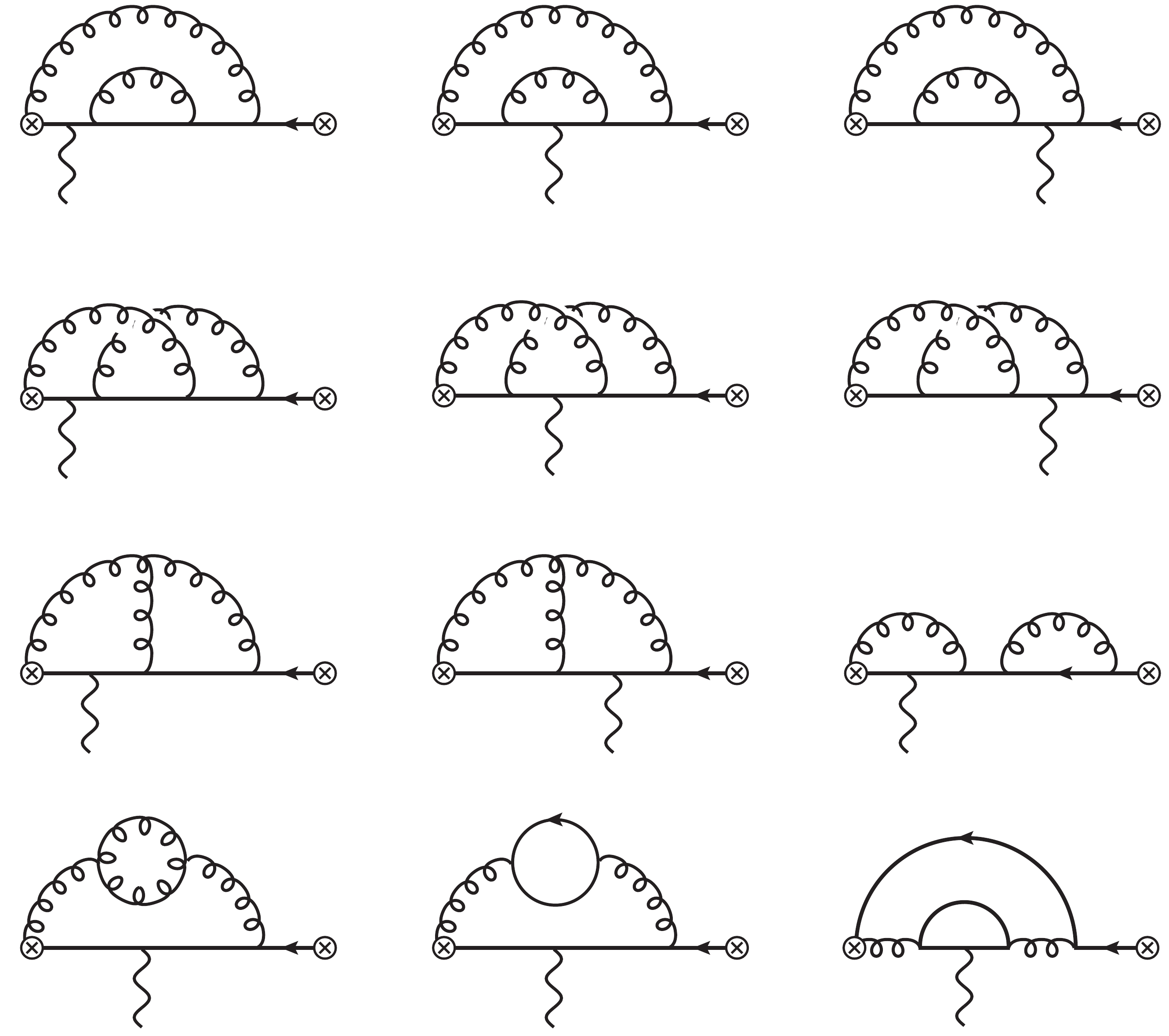}
\end{center}
\vspace{-2mm}
\caption{\label{fig:JG2Ldiags} 
Non-vanishing two-loop Feynman diagrams contributing to $J_G$ in light-cone gauge. Graphs with other photon attachments vanish, because they involve scaleless integrals. There are two  contributions from the last diagram with different orientations of fermion-number flow in the inner loop. Their sum vanishes because of Furry's theorem.}
\end{figure}

Figure~\ref{fig:JG2Ldiags} illustrates the non-vanishing two-loop Feynman diagrams contributing to $J_G$ in light-cone gauge. After performing simplifications of the Dirac, Lorentz and color algebras, these diagrams can be transformed into  linear combinations of scalar Feynman integrals belonging to one of five different integral topologies, each of which contains up to seven linearly independent squared propagators and up to two linear propagators. Mapping the Feynman integrals to specific integral topologies requires partial-fraction decompositions on linear propagators followed by shifts of the loop momenta. The five integral topologies can be cast into the form
\begin{equation}\label{2lgenintprop}
\begin{aligned}
  &\int {\rm d}^dl_1 \int{\rm d}^dl_2\,
   \frac{1}{{\cal D}_1^{a_1}\,{\cal D}_2^{a_2}\,{\cal D}_3^{a_3}\,{\cal D}_4^{a_4}\,{\cal D}_5^{a_5}\,
            {\cal D}_6^{a_6}\,{\cal D}_7^{a_7}\,{\cal D}_8^{a_8}\,{\cal D}_9^{a_9}\,
            {\cal D}_{10}^{a_{10}}\,{\cal D}_{11}^{a_{11}}\,{\cal D}_{12}^{a_{12}}} \\
  &= \mbox{}- \pi^d \left(-p^2\right)^{d-A_1} (\bar n\cdot p)^{-A_2}\,I_{\vec{a}}(d) \,,
\end{aligned}
\end{equation}
where $A_1=\sum_{i=1}^9 a_i$ and $A_2=\sum_{i=10}^{12} a_i$. The denominators ${\cal D}_n$ are defined as (omitting the ``$-i0$'' terms for brevity)
\begin{equation}
\begin{aligned}
   &{\cal D}_1 = -l_1^2 \,,
   & &{\cal D}_2 = -l_2^2 \,,
   & &{\cal D}_3 = -(l_1+l_2)^2 \,, \\
   &{\cal D}_4 = -(l_1+p)^2 \,, 
   & &{\cal D}_5 = -(l_2+p)^2 \,,
   & &{\cal D}_6 = -(l_1+l_2+p)^2 \,, \\
   &{\cal D}_7 = -(l_1+p-k)^2 \,, \quad
   & &{\cal D}_8 = -(l_2+p-k)^2 \,, \quad
   & &{\cal D}_9 = -(l_1+l_2+p-k)^2 \,, \\
   &{\cal D}_{10} = -{\bar n}\cdot l_1 \,,
   & &{\cal D}_{11} = -{\bar n}\cdot l_2 \,,
   & &{\cal D}_{12} = -{\bar n}\cdot (l_1+l_2) \,.
\end{aligned}
\end{equation}
The five integral topologies are given by restricting the propagator index in (\ref{2lgenintprop}) as follows:
\begin{equation}
\begin{aligned}
   &\text{topology 1:}\quad a_5=a_8=a_{12}=0 \,, &\quad
   &\text{topology 2:}\quad a_5=a_8=a_{11}=0 \,, \\
   &\text{topology 3:}\quad a_5=a_8=a_{10}=0 \,, &
   &\text{topology 4:}\quad a_3=a_8=a_{12}=0 \,, \\
   &\text{topology 5:}\quad a_3=a_6=a_{12}=0 \,. &
\end{aligned}
\end{equation}

We use the public program $\texttt{FIRE5}$ \cite{Smirnov:2014hma} to perform the integration-by-parts (IBP) reduction of the integrals in each of the five topologies. Besides the ten IBP identities in each topology, obtained by inserting differential operators in front of the integrands, additional linear algebraic identities can be derived by means of a kinematic constraint. The fact that the soft momentum $p_{s+}=(p-k)$ in (\ref{defJF}) is light-like leads to the relation
\begin{equation}
   p^\mu - k^\mu = \frac{p^2}{{\bar n}\cdot p}\,\frac{\bar n^\mu}{2} \,.
\end{equation}
By contracting both sides of this identity with the loop momenta $l_1^\mu$ and $l_2^\mu$ we derive the identities
\begin{equation}
\begin{aligned}
   {\bm a}_1^- - {\bm a}_7^- + \frac{p^2}{\bar n\cdot p}\,{\bm a}_{10}^- 
   &= 0 &\quad &\text{for topologies 1, 2, 4 and 5} \,, \\
   {\bm a}_1^- - {\bm a}_7^- - \frac{p^2}{\bar n\cdot p}\,({\bm a}_{11}^- - {\bm a}_{12}^-) 
   &= 0 &\quad &\text{for topology 3} \,, \\
\end{aligned}
\end{equation}
and
\begin{equation}
\begin{aligned}
   - {\bm a}_1^- + {\bm a}_3^- + {\bm a}_7^- - {\bm a}_9^- 
    + \frac{p^2}{\bar n\cdot p}\,{\bm a}_{11}^-
   &= 0 &\quad &\text{for topologies 1 and 3} \,, \\
   - {\bm a}_1^- + {\bm a}_3^- + {\bm a}_7^- - {\bm a}_9^- 
    - \frac{p^2}{\bar n\cdot p}\,({\bm a}_{10}^- 
    - {\bm a}_{12}^-) &= 0 &\quad &\text{for topology 2} \,, \\
   {\bm a}_2^- - {\bm a}_4^- - {\bm a}_5^- + {\bm a}_6^- + {\bm a}_7^- - {\bm a}_9^- 
    + \frac{p^2}{\bar n\cdot p}\,{\bm a}_{11}^- &= p^2 &\quad &\text{for topology 4} \,, \\
   {\bm a}_2^- - {\bm a}_8^- + \frac{p^2}{\bar n\cdot p}\,{\bm a}_{11}^- &= 0 &\quad 
    &\text{for topology 5} \,.
\end{aligned}
\end{equation}
Here the operator ${\bm a}_n^-$ lowers the index $a_n$ on the $n^{\rm th}$ denominator in (\ref{2lgenintprop}) by one unit. After implementing the above linear identities in $\texttt{FIRE5}$, the scalar Feynman integrals are reduced to ten master integrals (MIs), all of which can be remapped onto topology 4 by shifting the loop momenta. 

To compute the MIs analytically, we use a method inspired by \cite{Panzer:2014gra,vonManteuffel:2014qoa,vonManteuffel:2015gxa}, which has also been employed in the recent three-loop calculations of the soft and jet functions arising in the factorization theorems for the inclusive decays $\bar B\to X_s\gamma$ and $\bar B\to X_u\,\ell^-\bar\nu$ \cite{Bruser:2018rad,Bruser:2019yjk}. The basic strategy is to first map each MI in $d=4-2\epsilon$ spacetime dimensions to an analytically calculable quasi-finite integral in higher dimension $d=6-2\epsilon$ or $d=8-2\epsilon$, and then determine the linear relations between the MIs and the corresponding quasi-finite integrals by dimensional recurrence relations \cite{Tarasov:1996br,Lee:2009dh,Lee:2010wea} and IBP reduction. The quasi-finite integrals are free of divergences from the Feynman parameter integrations. They can be found by observing that raising the dimension by an even number decreases the degree of infrared divergences, while increasing appropriate propagator indices by integer amounts decreases the degree of ultraviolet divergences. The $\epsilon$ expansions of the quasi-finite integrals are linearly reducible and can be analytically evaluated by the powerful $\texttt{Maple}$ package $\texttt{HyperInt}$ \cite{Panzer:2014caa}. We use the public code $\texttt{LiteRed}$ \cite{Lee:2012cn,Lee:2013mka} to determine the dimensional recurrence relations. After further IBP reduction, we construct the linear relations between the MIs in $d+2$ and $d$ dimension in the form $\vec{I}(d+2)={\bm A}(d)\cdot \vec{I}(d)$. This allows us to build linear relations between the MIs in $d=4-2\epsilon$ and the corresponding quasi-finite integrals in higher dimension. Then the analytical results of the MIs are obtained by solving the linear equations.

With the integrals at hand, the jet function $J_G$ is obtained by evaluating the Feynman diagrams shown in Figure~\ref{fig:JG2Ldiags}. Adding to this result the contribution from $J_A$, we obtain for the bare jet function at two-loop order
\begin{equation}\label{Jbare}
\begin{aligned}
   J^{(0)}(p^2) 
   &= 1 + \frac{Z_\alpha\alpha_s}{4\pi} \left( \frac{-p^2}{\mu^2} \right)^{-\epsilon}
    C_F\,e^{\epsilon\gamma_E}\,\frac{\Gamma(1+\epsilon)\,\Gamma^2(-\epsilon)}{\Gamma(2-2\epsilon)}\,
    (2-4\epsilon-\epsilon^2) \\
   &\quad\mbox{}+ \left( \frac{Z_\alpha\alpha_s}{4\pi} \right)^2
    \left( \frac{-p^2}{\mu^2} \right)^{-2\epsilon}
    C_F\,\big( C_F K_F + C_A K_A + T_F\,n_f K_{n_f} \big)
    + {\cal O}(\alpha_s^3) \,,
\end{aligned}
\end{equation}
where the two-loop coefficients are given by
\begin{equation}
\begin{aligned}
   K_F &= \frac{2}{\epsilon^4} + \frac{1}{\epsilon^2} \left( - 2 - \frac{\pi^2}{3} \right)
    + \frac{1}{\epsilon} \left( - 4 - \frac{\pi^2}{2} - \frac{46\zeta_3}{3} \right)
    - \frac{13}{2} - \frac{\pi^2}{6} - 39\zeta_3 + \frac{\pi^4}{5} + {\cal O}(\epsilon) \,, \\
   K_A &= \frac{11}{6\epsilon^3} + \frac{1}{\epsilon^2} \left( \frac{67}{18} - \frac{\pi^2}{6} \right)
    + \frac{1}{\epsilon} \left( \frac{103}{27} - \frac{11\pi^2}{36} - 7\zeta_3 \right)\!
    - \frac{695}{162} - \frac{103\pi^2}{108} - \frac{14\zeta_3}{9} - \frac{43\pi^4}{180} 
    + \hspace{-0.5mm} {\cal O}(\epsilon) \hspace{-0.5mm} \,, \\
   K_{n_f} &= - \frac{2}{3\epsilon^3} - \frac{10}{9\epsilon^2} 
    + \frac{1}{\epsilon} \left( - \frac{20}{27} + \frac{\pi^2}{9} \right)
    + \frac{230}{81} + \frac{5\pi^2}{27} + \frac{64\zeta_3}{9} + {\cal O}(\epsilon) \,.
\end{aligned}
\end{equation}
In Appendix~\ref{app:A} we present our results for the two jet functions $J_A$ and $J_G$ separately, including terms up to and including ${\cal O}(\epsilon^2)$.

Given the above result for the bare jet function in momentum space, it is straightforward to derive the corresponding expression for the bare jet function in the dual space. The first relation in (\ref{trafo}) implies that this function can be obtained from (\ref{Jbare}) by means of the replacement
\begin{equation}
   \left( \frac{-p^2}{\mu^2} \right)^{-n\epsilon}
    \!\to\, \frac{\Gamma(1-n\epsilon)}{\Gamma(1+n\epsilon)}
    \left( \frac{-p^2}{\mu^2} \right)^{-n\epsilon} .
\end{equation}

\section{Renormalization of the jet function}
\label{sec:4}

Because of the relative simplicity of the anomalous dimension in (\ref{gammaJ2loop}), it is most convenient to perform the renormalization of the jet function in the dual space. Given the anomalous dimension, one can construct the renormalization factor $Z_J^{\rm dual}$ in the dual space, defined in analogy with (\ref{Jrenorm}), using a general relation derived in \cite{Becher:2009cu}. It is based on a formal solution of relation (\ref{neweq}) and applies to Sudakov problems, in which the anomalous dimension contains an explicit dependence on $\ln\mu$. We obtain
\begin{equation}
   Z_J^{\rm dual}(p^2,x p^2;\mu) \\
   = \exp_\otimes\! \Bigg[ 
    \int\limits_0^{\alpha_s(\mu)}\!\!\frac{d\alpha}{2\epsilon\alpha-\beta(\alpha)}\,
    \bigg[ \gamma_J^{\rm dual}(y_i\hspace{0.3mm}p^2,y_i'\hspace{0.3mm}p^2;\mu) 
    - \delta(y_i-y_i') \int\limits_0^\alpha d\alpha'\,
    \frac{2\Gamma_{\rm cusp}(\alpha')}{2\epsilon\alpha'-\beta(\alpha')} \bigg] \Bigg] .
\end{equation}
This exact solution must be expanded in powers of $\alpha_s$ at fixed $\epsilon$ to generate the renormalization factor in the $\overline{\rm MS}$ scheme. Note that the anomalous dimension $\gamma_J^{\rm dual}(y_i\hspace{0.3mm}p^2,y_i'\hspace{0.3mm}p^2;\mu)$ in the exponent must be evaluated with $\alpha_s(\mu)$ replaced by $\alpha$. The expression on the right-hand side of the equation is a generalized matrix exponential. The symbol $\otimes$ means that, when the exponential is expanded in a power series, one must integrate over all variables $y_i$ and $y_i'$  except for the first and the last one, which must be set equal to 1 and $x$, respectively. This yields
\begin{equation}
\begin{aligned}
   Z_J^{\rm dual}(p^2,x p^2;\mu) 
   &= \delta(1-x) + \!\!\int\limits_0^{\alpha_s(\mu)}\!\!
    \frac{d\alpha}{2\epsilon\alpha-\beta(\alpha)}\,\bigg[ \gamma_J^{\rm dual}(p^2,x p^2;\mu) 
    - \delta(1-x) \int\limits_0^\alpha d\alpha'\,
    \frac{2\Gamma_{\rm cusp}(\alpha')}{2\epsilon\alpha'-\beta(\alpha')} \bigg] \\
   &\mbox{}+ \frac{1}{2} \int_0^\infty\!dy\!
    \int\limits_0^{\alpha_s(\mu)}\! \frac{d\alpha}{2\epsilon\alpha-\beta(\alpha)}\,
    \bigg[ \gamma_J^{\rm dual}(p^2,y p^2;\mu) 
    - \delta(1-y) \int\limits_0^\alpha d\alpha'\,
    \frac{2\Gamma_{\rm cusp}(\alpha')}{2\epsilon\alpha'-\beta(\alpha')} \bigg] \\
   &\quad\times \!\int\limits_0^{\alpha_s(\mu)}\! \frac{d\alpha}{2\epsilon\alpha-\beta(\alpha)}\,
    \bigg[ \gamma_J^{\rm dual}(y p^2,x p^2;\mu) 
    - \delta(y-x) \int\limits_0^\alpha d\alpha'\,
    \frac{2\Gamma_{\rm cusp}(\alpha')}{2\epsilon\alpha'-\beta(\alpha')} \bigg] 
    + \dots \,.
\end{aligned}
\end{equation} 
At two-loop order, we obtain in this way
\begin{equation}\label{ZJdual}
\begin{aligned}
   Z_J^{\rm dual}(p^2,x p^2;\mu)
   &= \Bigg\{ 1 + \frac{\alpha_s}{4\pi}\,\bigg[ - \frac{\Gamma_0}{2\epsilon^2} 
    + \frac{\Gamma_0 \hat L_p-\gamma_0'}{2\epsilon} \bigg] 
    + \left( \frac{\alpha_s}{4\pi} \right)^2 \bigg[
    \frac{\Gamma_0^2}{8\epsilon^4} - \frac{\Gamma_0}{4\epsilon^3} 
    \left( \Gamma_0 \hat L_p - \gamma_0' - \frac32\,\beta_0 \right) \\
   &\hspace{1.1cm}\mbox{}+ \frac{1}{8\epsilon^2} \left( \Gamma_0 \hat L_p-\gamma_0' \right) 
    \left( \Gamma_0 \hat L_p-\gamma_0' - 2\beta_0 \right) - \frac{\Gamma_1}{8\epsilon^2}
    + \frac{\Gamma_1 \hat L_p-\gamma_1'}{4\epsilon} \bigg] \Bigg\}\,\delta(1-x) \\
   &\quad\mbox{}+ \frac{C_F}{\epsilon} \left( \frac{\alpha_s}{4\pi} \right)^2
    \frac{\theta(1-x)}{1-x}\,h(x) + {\cal O}(\alpha_s^3) \,,
 \end{aligned}
\end{equation}
where $\hat L_p=\ln(-p^2 e^{-2\gamma_E}/\mu^2)$. The relevant expansion coefficients $\Gamma_n$ and $\gamma'_n$ of the anomalous dimensions are listed in Appendix~\ref{app:B}. Applying this result to the bare jet function in the dual space, we find that indeed all $1/\epsilon$ poles cancel out. This provides a non-trivial consistency check on the two-loop results for $\gamma_\eta(\alpha_s)$ in (\ref{gammaprela}) and $h(x)$ in (\ref{hdef}), which were obtained in \cite{Braun:2019wyx}. The result for the renormalized dual jet function reads
\begin{equation}\label{j2loop}
   j(p^2,\mu) = 1 + \frac{C_F\alpha_s}{4\pi} \left[ \bigg( \hat L_p^2 - 1 - \frac{\pi^2}{6} \bigg) 
    + \frac{\alpha_s}{4\pi}\,\big( C_F k_F^{\rm dual} + C_A k_A^{\rm dual} 
    + T_F\,n_f\hspace{0.3mm} k_{n_f}^{\rm dual} \big) \right] 
    + {\cal O}(\alpha_s^3) \,,
\end{equation}
where
\begin{equation}
\begin{aligned}
   k_F^{\rm dual} &= \frac{\hat L_p^4}{2} - \left( 1 + \frac{\pi^2}{6} \right) \hat L_p^2 
    + \left( \pi^2 - 4\zeta_3 \right) \hat L_p + \frac32 - \frac{\pi^2}{3} - 39\zeta_3
    + \frac{119\pi^4}{360} \,, \\
   k_A^{\rm dual} &= - \frac{11}{9}\,\hat L_p^3 
    + \left( \frac{67}{9} - \frac{\pi^2}{3} \right) \hat L_p^2 
    - \left( \frac{305}{27} - 14\zeta_3 \right) \hat L_p + \frac{493}{162} 
    - \frac{103\pi^2}{108} + \frac{184\zeta_3}{9} - \frac{43\pi^4}{180} \,, \\
   k_{n_f}^{\rm dual} &= \frac{4}{9}\,\hat L_p^3 - \frac{20}{9}\,\hat L_p^2 
    + \frac{76}{27}\,\hat L_p + \frac{14}{81} + \frac{5\pi^2}{27} - \frac{8\zeta_3}{9} \,.
\end{aligned}
\end{equation}

Given the above result, we can obtain the jet function in momentum space by applying the second integral transformation in (\ref{trafo}). It follows that we must perform the replacements
\begin{equation}
   \hat L_p\to L_p \,, \qquad
   \hat L_p^2\to L_p^2 \,, \qquad
   \hat L_p^3\to L_p^3 + 4\zeta_3 \,, \qquad
   \hat L_p^4\to L_p^4 + 16\zeta_3 L_p \,,
\end{equation}
where now $L_p=\ln(-p^2/\mu^2)$. This gives the final result
\begin{equation}\label{J2loop}
   J(p^2,\mu) = 1 + \frac{C_F\alpha_s}{4\pi} \left[ \bigg( L_p^2 - 1 - \frac{\pi^2}{6} \bigg) 
   + \frac{\alpha_s}{4\pi}\,\big( C_F k_F + C_A k_A + T_F\,n_f\hspace{0.3mm} k_{n_f} \big) \right] 
   + {\cal O}(\alpha_s^3) \,,
\end{equation}
with
\begin{equation}
\begin{aligned}
   k_F &= \frac{L_p^4}{2} - \left( 1 + \frac{\pi^2}{6} \right) L_p^2 
    + \left( \pi^2 + 4\zeta_3 \right) L_p + \frac32 - \frac{\pi^2}{3} - 39\zeta_3
    + \frac{119\pi^4}{360} \,, \\
   k_A &= - \frac{11}{9}\,L_p^3 + \left( \frac{67}{9} - \frac{\pi^2}{3} \right) L_p^2 
    - \left( \frac{305}{27} - 14\zeta_3 \right) L_p + \frac{493}{162} 
    - \frac{103\pi^2}{108} + \frac{140\zeta_3}{9} - \frac{43\pi^4}{180} \,, \\
   k_{n_f} &= \frac{4}{9}\,L_p^3 - \frac{20}{9}\,L_p^2 + \frac{76}{27}\,L_p 
    + \frac{14}{81} + \frac{5\pi^2}{27} + \frac{8\zeta_3}{9} \,.
\end{aligned}
\end{equation}
The two-loop expressions (\ref{j2loop}) and (\ref{J2loop}) for the radiative jet functions in the dual space and in momentum space are important new results obtained in this paper.

We have derived the two-loop result (\ref{J2loop}) for $J(p^2)$ in a second, totally independent way by performing a two-loop calculation of the hard matching coefficients $H_2(z)$ and $H_3$ in the factorization formula for the $h\to\gamma\gamma$ decay amplitude derived in \cite{Liu:2019oav}. Further details about this rather difficult calculation will be presented elsewhere. In the limit where $z\to 0$, these coefficients were shown to obey the refactorization formula
\begin{equation}\label{refac}
   \lim_{z\to 0}\,H_2(z) = - \frac{H_3}{z}\,\Big[ J(z M_h^2) + {\cal O}(z) \Big] \,.
\end{equation}
The two-loop result we have obtained for the bare jet function from this relation is in complete agreement with the expression given above. This not only confirms our result for the jet function but also provides a non-trivial test of the refactorization formula (\ref{refac}).

\section{\boldmath Phenomenological impact of the two-loop corrections}
\label{sec:5}

The two-loop jet function we have calculated in this paper is the last missing ingredient for a calculation of the leading-power contributions to the $B^-\to\gamma\ell^-\bar\nu$ decay amplitude at NNLO in RG-improved perturbation theory. This is important, since this process provides the most direct information about the properties of the $B$-meson LCDA \cite{Beneke:2011nf}. The potential impact of power corrections in $\Lambda_{\rm QCD}/m_b$ has been studied in \cite{Wang:2016qii,Wang:2018wfj,Beneke:2018wjp}. 

In order to estimate the potential impact of the two-loop corrections to the jet function, we recall that, as shown in (\ref{QCDF}), the decay amplitude is proportional to the convolution
\begin{equation}
   J\otimes\phi
   = \int_0^\infty\!\frac{d\omega}{\omega}\,J(-2E_\gamma\omega,\mu)\,\phi_+^B(\omega,\mu) \,,
\end{equation}
where $E_\gamma$ is the photon energy in the $B$-meson rest frame. For the purposes of illustration, we fix the renormalization scale at a value $\mu=\mu_j\approx 1.5$\,GeV, corresponding to a typical matching scale for the jet function. The LCDA naturally lives at a lower scale $\mu_0$ of order 1\,GeV, but since in practice there is no large scale hierarchy, we will not resum logarithms of the ratio $\mu_j/\mu_0$. Following \cite{Beneke:2011nf}, we define the hadronic matrix elements
\begin{equation}
   \frac{1}{\lambda_B(\mu)} = \int_0^\infty\!\frac{d\omega}{\omega}\,\phi_+^B(\omega,\mu) \,, \qquad
   \sigma_n(\mu) = \lambda_B(\mu) \int_0^\infty\!\frac{d\omega}{\omega}\,
    \ln^n\frac{\mu_0}{\omega}\,\phi_+^B(\omega,\mu) \,,
\end{equation}
where $\mu_0=1$\,GeV is a fixed reference scale, which is part of the definition of the logarithmic moments. For simplicity, we choose the matching scale $\mu_j$ such that $\ln(2E_\gamma\mu_0/\mu_j^2)=0$. We then obtain at two-loop order (with $n_f=4$ light quark flavors)
\begin{equation}
\begin{aligned}
   J\otimes\phi
   &= \frac{1}{\lambda_B}\,\bigg\{ 1 + \frac{\alpha_s(\mu_j)}{0.35}\, 
    \big( 3.71\sigma_2 - 9.82 \big)\cdot 10^{-2} \\
   &\hspace{1.73cm}\mbox{}+ \left[ \frac{\alpha_s(\mu_j)}{0.35} \right]^2 
    \big( 0.07\sigma_4 + 0.29\sigma_3 + 0.46\sigma_2 - 4.32\sigma_1 - 5.03 \big)\cdot 10^{-2} 
    \bigg\} \,,
\end{aligned}
\end{equation}
where all hadronic parameters are defined at the scale $\mu_j$. We have chosen $\alpha_s(\mu_j)=0.35$ as a reference value for the strong coupling, which corresponds to $\mu_j\approx 1.5$\,GeV. On general grounds, one expects the logarithmic moments $\sigma_n$ to be of ${\cal O}(1)$, even though in concrete models for $\phi_+^B(\omega)$ such as the simplest exponential model \cite{Grozin:1996pq} the higher moments defined with $\mu_0=1$\,GeV tend to take on much larger values. We observe that the two-loop corrections are not significantly smaller than the corrections arising at one-loop order and hence should be included in future analyses. For example, using the central values of the default choices $\sigma_1=1.5\pm 1$ and $\sigma_2=3\pm 2$ adopted in \cite{Beneke:2011nf}, one finds
\begin{equation}
   J\otimes\phi
   \approx \frac{1}{\lambda_B}\,\Big[ 1 + 1.3\,\%\,\big|_{\alpha_s} 
    + \big( - 10.1 + 0.29\sigma_3 + 0.069\sigma_4 \big) \%\,\big|_{\alpha_s^2}\, \Big] \,.
\end{equation}
Note that the smallness of the one-loop correction term is linked with the particular choice of $\sigma_2$. We leave a complete analysis of the $B^-\to\gamma\ell^-\bar\nu$ decay rate including higher-order perturbative corrections for future work.

\section{Conclusions}

We have presented a detailed study of the radiative jet function $J(p^2)$ defined in (\ref{defJF}), which plays a central role in the theory of factorization at subleading power in scale ratios. This object appears in factorization theorems for important exclusive processes such as the rare $B$-meson decay $B^-\to\gamma\ell^-\bar\nu$ and the contributions to the radiative Higgs-boson decay $h\to\gamma\gamma$ mediated by light-quark loops. In the first case, in particular, a precise knowledge of the jet function is a prerequisite for extracting accurate information about the $B$-meson LCDA from experimental data in the region of high photon energy. The $B$-meson LCDA itself is a central quantity in the theory of QCD factorization applied to exclusive decays of $B$ mesons \cite{Beneke:1999br,Beneke:2000ry}.

We have calculated the radiative jet function at two-loop order both in momentum space and in a dual space, in which its RG evolution equation takes on a particularly simple form. We have further derived the anomalous dimensions for the jet functions in momentum and the dual space, including for the first time the so-far unknown two-loop contributions not controlled by the cusp anomalous dimension. Our derivation of these terms has relied on a recent calculation of the corresponding contributions to the anomalous dimension of the $B$-meson LCDA in the dual space \cite{Braun:2019wyx}. We find that the quantity $\gamma'(\alpha_s)$ in (\ref{gammaJ}) obeys an unexpected relation with the anomalous dimension of the Drell-Yan soft function \cite{Becher:2007ty}, which deserves further study. Finally, we have obtained analytic solutions to the two-loop RG evolution equations of the radiative jet function in both spaces. The technique we used for obtaining these solutions is general and can be applied to other radiative jet functions as well.

The results presented in this paper will play an important role in the renormalization of the factorization theorem for the light-quark induced $h\to\gamma\gamma$ decay amplitude derived in \cite{Liu:2019oav}. Using the evolution equations of the radiative jet function derived here and of a hard function well known in SCET, we will be able to derive the two-loop evolution equation satisfied by the soft-quark soft function \cite{Liu:2020eqe}, another central object in the theory of factorization beyond the leading power.

\subsubsection*{Acknowledgements}
One of us (M.N.) is grateful to Thomas Becher, Guido Bell, Martin Beneke, Volodya Braun and Thorsten Feldmann for useful discussions. This research has been supported by the Cluster of Excellence PRISMA$^+$\! (project ID 39083149), funded by the German Research Foundation (DFG), and under grant 05H18UMCA1 of the German Federal Ministry for Education and Research (BMBF). The research of Z.L.L.\ is supported by the U.S.\ Department of Energy under Contract No.~DE-AC52-06NA25396, the LANL/LDRD program and within the framework of the TMD Topical Collaboration.

\newpage
\begin{appendix}
\renewcommand{\theequation}{A.\arabic{equation}}
\setcounter{equation}{0}

\section{\boldmath Evolution with a general boundary condition}
\label{app:C}

For the case of downward scale evolution, for which $\mu<\mu_j$ and hence $a_\Gamma(\mu_j,\mu)>0$, it is possible to evaluate the integral over $t$ in (\ref{Jsol}) in closed form using the theorem of residues, adopting a technique developed in \cite{Lee:2005gza} for the case of the $B$-meson LCDA. The integrand contains single poles in the complex $t$-plane located at $t=in$ and $t=-i[n-a_\Gamma(\mu_j,\mu)]$, where $n\in\mathbb{N}$ is a positive integer. Expressing the Fourier image $\tilde J(t,\mu_j)$ in terms of the original jet function using
\begin{equation}\label{inverseFT}
   \tilde J(t,\mu) = \int_0^\infty\!\frac{dx}{x}\,J(x p^2,\mu)
    \left( \frac{-x p^2}{\mu^2} \right)^{-it} ,   
\end{equation}
we see that for $x<1$ ($x>1$) the contour can be closed in the upper (lower) half plane, and one then needs to perform the infinite sum over the residues of the poles. Let us assume that $a_\Gamma(\mu_j,\mu)<1$, such that all poles in the second series lie in the lower half plane. This condition is always satisfied in practical calculations. At leading order, it is equivalent to the statement that $\alpha_s(\mu)/\alpha_s(\mu_j)>\exp\big(\frac{2\beta_0}{\Gamma_0}\big)\approx 23$, where the numerical value refers to the case of four light quark flavors. We then obtain
\begin{equation}\label{JsolFourier}
\begin{aligned}
   J(p^2,\mu) 
   &= \exp\Big[ -2S(\mu_j,\mu) - a_{\gamma'}(\mu_j,\mu) \Big]\,
    e^{-2\gamma_E\,a}\,\frac{\Gamma(2-a)}{\Gamma(a)}\,\bigg( \frac{-p^2}{\mu_j^2} \bigg)^a \\
   &\quad\times\int_0^1\!dx\,\Big[ J(x p^2,\mu_j) + x^{-a} J(p^2/x,\mu_j) \Big]\,\,
    {}_2F_1(1-a,2-a;2;x) \,,
\end{aligned}
\end{equation}
where $a\equiv a_\Gamma(\mu_j,\mu)$. This elegant formula relates the jet function at the scale $\mu<\mu_j$ to the jet function defined at the matching scale $\mu_j$, irrespective of what the initial condition is. We stress that this formula cannot be applied to the case of upward evolution ($\mu>\mu_j$). In the limit $x\to 1$ and for $a<\frac12$, the hypergeometric function behaves like ${}_2F_1(1-a,2-a;2;x)\sim(1-x)^{-1+2a}$, which generates a non-integrable singularity if $a<0$.

\renewcommand{\theequation}{B.\arabic{equation}}
\setcounter{equation}{0}

\section{\boldmath Two-loop results for the functions $J_A$ and $J_G$}
\label{app:A}

Here we present our results for the jet functions $J_A$ and $J_G$ defined in (\ref{JAJGdef}), keeping terms of higher order in $\epsilon$. We find
\begin{equation}
\begin{aligned}
   J_A^{(0)}(p^2) 
   &= 1 + \frac{Z_\alpha\alpha_s}{4\pi} \left( \frac{-p^2}{\mu^2} \right)^{-\epsilon}
    C_F\,e^{\epsilon\gamma_E}\,\frac{\Gamma(1+\epsilon)\,\Gamma^2(-\epsilon)}{\Gamma(2-2\epsilon)}\,
    (4-5\epsilon+\epsilon^2) \\
   &\quad\mbox{}+ \left( \frac{Z_\alpha\alpha_s}{4\pi} \right)^2
    \left( \frac{-p^2}{\mu^2} \right)^{-2\epsilon}
    C_F\,\big( C_F K_F^A + C_A K_A^A + T_F\,n_f K_{n_f}^A \big)
    + {\cal O}(\alpha_s^3) \,, \\
   J_G^{(0)}(p^2) 
   &= 1 + \frac{Z_\alpha\alpha_s}{4\pi} \left( \frac{-p^2}{\mu^2} \right)^{-\epsilon}
    C_F\,e^{\epsilon\gamma_E}\,\frac{\Gamma(1+\epsilon)\,\Gamma^2(-\epsilon)}{\Gamma(2-2\epsilon)}\,
    (-2+\epsilon-2\epsilon^2) \\
   &\quad\mbox{}+ \left( \frac{Z_\alpha\alpha_s}{4\pi} \right)^2
    \left( \frac{-p^2}{\mu^2} \right)^{-2\epsilon}
    C_F\,\big( C_F K_F^G + C_A K_A^G + T_F\,n_f K_{n_f}^G \big)
    + {\cal O}(\alpha_s^3) \,. 
\end{aligned}
\end{equation}
For the two-loop coefficients in these expressions, we obtain up to ${\cal O}(\epsilon^2)$ the following results:
{\small
\begin{equation}
\begin{aligned}
   K_F^A &= \frac{8}{\epsilon^4} + \frac{12}{\epsilon^3} + \frac{65}{2\epsilon^2}
    + \frac{1}{\epsilon} \left( \frac{311}{4} - \pi^2 -\frac{124\zeta_3}{3} \right)
    + \left( \frac{1437}{8} - \frac{41\pi^2}{12} - 86\zeta_3 + \frac{17\pi^4}{90} \right) \\
   &\quad\mbox{}+ \left( \frac{6479}{16} - \frac{215\pi^2}{24} - \frac{878\zeta_3}{3}
    - \frac{\pi^4}{15} - \frac{4\pi^2\zeta_3}{3} + \frac{164\zeta_5}{5} \right) \epsilon \\
   &\quad\mbox{}+ \left( \frac{28589}{32} - \frac{351\pi^2}{16} - \frac{2344\zeta_3}{3}
    - \frac{95\pi^4}{48} + \frac{35\pi^2\zeta_3}{3} - \frac{474\zeta_5}{5}
    + \frac{491\pi^6}{1620} + \frac{1654\zeta_3^2}{9} \right) \epsilon^2 \,, \\
   K_A^A &= \frac{11}{3\epsilon^3} 
    + \frac{1}{\epsilon^2} \left( \frac{233}{18} - \frac{\pi^2}{3} \right)
    + \frac{1}{\epsilon} \left( \frac{4541}{108} - \frac{11\pi^2}{18} - 20\zeta_3 \right) 
    + \left( \frac{86393}{648} - \frac{197\pi^2}{108} - \frac{514\zeta_3}{9}
    - \frac{19\pi^4}{60} \right) \\
   &\quad\mbox{}+ \left( \frac{1605689}{3888} - \frac{4109\pi^2}{648} - \frac{4646\zeta_3}{27}
    - \frac{317\pi^4}{360} + \frac{59\pi^2\zeta_3}{9} - 132\zeta_5 \right) \epsilon \\
   &\quad\mbox{}+ \left( \frac{29451137}{23328} - \frac{81209\pi^2}{3888} - \frac{43780\zeta_3}{81}
    - \frac{967\pi^4}{432} + \frac{257\pi^2\zeta_3}{27} - \frac{4882\zeta_5}{15}
    - \frac{421\pi^6}{2835} + \frac{403\zeta_3^2}{3} \right) \epsilon^2 \,, \\
   K_{n_f}^A &= - \frac{4}{3\epsilon^3} - \frac{38}{9\epsilon^2} 
    + \frac{1}{\epsilon} \left( - \frac{373}{27} + \frac{2\pi^2}{9} \right)
    + \left( - \frac{7081}{162} + \frac{19\pi^2}{27} + \frac{128\zeta_3}{9} \right) \\
   &\quad\mbox{}+ \left( - \frac{131761}{972} + \frac{373\pi^2}{162} + \frac{1216\zeta_3}{27}
    + \frac{19\pi^4}{90} \right) \epsilon \\
   &\quad\mbox{}+ \left( - \frac{2422201}{5832} + \frac{7081\pi^2}{972} + \frac{11936\zeta_3}{81}
    + \frac{361\pi^4}{540} - \frac{64\pi^2\zeta_3}{27} + \frac{1088\zeta_5}{15} \right) 
    \epsilon^2 \,,
\end{aligned}
\end{equation}
}
and
{\small
\begin{equation}
\begin{aligned}
   K_F^G &= - \frac{6}{\epsilon ^4} - \frac{12}{\epsilon ^3} 
    + \frac{1}{\epsilon^2} \left( - \frac{69}{2} - \frac{\pi^2}{3} \right)
    + \frac{1}{\epsilon} \left( - \frac{327}{4} + \frac{\pi^2}{2} + 26\zeta _3 \right) 
    + \left( - \frac{1489}{8} + \frac{13\pi^2}{4} + 47\zeta_3
    + \frac{\pi^4}{90} \right) \\
   &\quad\mbox{}+ \left( - \frac{6611}{16} + \frac{77\pi^2}{8} + 149\zeta_3
    - \frac{2\pi^4}{3} + \frac{38\pi^2\zeta_3}{9} + \frac{602\zeta_5}{5} \right) \epsilon \\
   &\quad\mbox{}+ \left( - \frac{28665}{32} + \frac{1201\pi^2}{48} + 188\zeta_3
    - \frac{431\pi^4}{720} + \frac{19\pi^2\zeta_3}{6} - \frac{291\zeta_5}{5}
    + \frac{3977\pi^6}{11340} + 47\zeta_3^2 \right) \epsilon^2 \,, \\
   K_A^G &= - \frac{11}{6\epsilon^3} 
    + \frac{1}{\epsilon^2} \left( - \frac{83}{9} + \frac{\pi^2}{6} \right)
    + \frac{1}{\epsilon} \left( - \frac{4129}{108} + \frac{11\pi^2}{36} + 13\zeta_3 \right) 
    + \left( - \frac{89173}{648} + \frac{47\pi^2}{54} + \frac{500\zeta_3}{9}
    + \frac{7\pi^4}{90} \right) \\
   &\quad\mbox{}+ \left( - \frac{1775893}{3888} + \frac{3049\pi^2}{648} + \frac{5788\zeta_3}{27}
    + \frac{133\pi^4}{144} - \frac{65\pi^2\zeta_3}{18} + \zeta_5 \right) \epsilon \\
   &\quad\mbox{}+ \left( - \frac{33912061}{23328} + \frac{74917\pi^2}{3888} + \frac{66242\zeta_3}{81}
    + \frac{643\pi^4}{216} - \frac{304\pi^2\zeta_3}{27} + \frac{4466\zeta_5}{15}
    - \frac{245\pi^6}{1296} - \frac{467\zeta_3^2}{3} \right) \epsilon^2 \,, \\ \nonumber
\end{aligned}
\end{equation}
\begin{equation}
\begin{aligned}
   K_{n_f}^G &= \frac{2}{3\epsilon^3} + \frac{28}{9\epsilon^2}
    + \frac{1}{\epsilon} \left( \frac{353}{27} - \frac{\pi^2}{9} \right)
    + \left( \frac{7541}{162} - \frac{14\pi^2}{27} - \frac{64\zeta_3}{9} \right) \\
   &\quad\mbox{}+ \left( \frac{150125}{972} - \frac{353\pi^2}{162} - \frac{896\zeta_3}{27}
    - \frac{19\pi^4}{180} \right) \epsilon \\
   &\quad\mbox{}+ \left( \frac{2877653}{5832} - \frac{7541\pi^2}{972} - \frac{11296\zeta_3}{81}
    - \frac{133\pi^4}{270} + \frac{32\pi^2\zeta_3}{27} - \frac{544\zeta_5}{15} \right) 
    \epsilon^2 \,. \hspace{4.35cm}
\end{aligned}
\end{equation}
}

\renewcommand{\theequation}{C.\arabic{equation}}
\setcounter{equation}{0}

\section{Anomalous dimensions}
\label{app:B}

Here we list expressions for the relevant anomalous dimensions up to two-loop order. We define the perturbative expansion coefficients via 
\begin{equation}
   \Gamma_{\rm cusp}(\alpha_s) 
   = \Gamma_0\,\frac{\alpha_s}{4\pi} + \Gamma_1 \left( \frac{\alpha_s}{4\pi} \right)^2 + \dots \,, 
\end{equation}
and similarly for all other anomalous dimensions. The coefficients needed in (\ref{ZJdual}) are
\begin{equation}
\begin{aligned}
   \Gamma_0 &= 4 C_F \,, \\
   \Gamma_1 &= 4 C_F \left[ \left( \frac{67}{9} - \frac{\pi^2}{3} \right) C_A
    - \frac{20}{9}\,T_F\,n_f \right] , \\
   \gamma'_0 &= 0 \,, \\
   \gamma'_1 &= C_F \left[ C_A \bigg( \frac{808}{27} - \frac{11\pi^2}{9} - 28\zeta_3 \bigg) 
    - T_F\,n_f \bigg( \frac{224}{27} - \frac{4\pi^2}{9} \bigg) \right] \,.
\end{aligned}
\end{equation}
The coefficient of the anomalous dimensions in (\ref{gprela}) are
\begin{equation}
\begin{aligned}
   \gamma_{\eta,0} &= -2 C_F \,, \\
   \gamma_{\eta,1} &= C_F \left[ C_F \left( - 4 + \frac{14\pi^2}{3} - 24\zeta_3 \right)
    + C_A \left( \frac{254}{27} - \frac{55\pi^2}{18} - 6\zeta_3 \right) 
    + T_F\,n_f \left( - \frac{64}{27} + \frac{10\pi^2}{9} \right) \right] , \\
   \gamma_{H,0} &= -5 C_F \,, \\
   \gamma_{H,1} &= C_F \left[ C_F \left( - \frac32 + 2\pi^2 - 24\zeta_3 \right)
    + C_A \left( - \frac{1549}{54} - \frac{7\pi^2}{6} + 22\zeta_3 \right) 
    + T_F\,n_f \left( \frac{250}{27} + \frac{2\pi^2}{3} \right) \right] , \\
   \gamma_{F,0} &= -3 C_F \,, \\
   \gamma_{F,1} &= C_F \left[ C_F \left( \frac52 - \frac{8\pi^2}{3} \right)
    + C_A \left( - \frac{49}{6} + \frac{2\pi^2}{3} \right) 
    + \frac{10}{3}\,T_F\,n_f \right] .
\end{aligned}
\end{equation}

\end{appendix}

\end{document}